\documentclass[preprint,showpacs,showkeys]{revtex4}%
\usepackage{amsfonts}
\usepackage{amsmath}
\usepackage{amssymb}
\usepackage{graphicx}
\usepackage[english]{babel}%
\providecommand{\U}[1]{\protect\rule{.1in}{.1in}}

\begin{document}
\title{Constraints on extra dimensions from atomic spectroscopy}
\author{F. Dahia}
\affiliation{{\footnotesize Department of Physics, Universidade Federal da Para\'{\i}ba,
Jo\~{a}o Pessoa - PB, Brazil}}
\author{A. S. Lemos }
\affiliation{{\footnotesize Department of Physics, Universidade Federal da Para\'{\i}ba,
Jo\~{a}o Pessoa - PB, Brazil}}

\begin{abstract}
We consider a hydrogen atom confined in a thick brane embedded in a
higher-dimensional space. Due to effects of the extra dimensions, the
gravitational potential is amplified in distances smaller than the size of the
supplementary space, in comparison with the Newtonian potential. Studying the
influence of the gravitational interaction modified by the extra dimensions on
the energy levels of the hydrogen atom, we find independent constraints for
the higher-dimensional Planck mass in terms of the thickness of the brane by
using accurate measurements of atomic transition frequencies. The constraints
are very stringent for narrow branes.

\end{abstract}
\maketitle

\section{Introduction}

\qquad Since the beginning of the past century, theories of higher-dimensional
spacetime have attracted attention of physicists mainly motivated by the
Kaluza-Klein theory, which postulated the existence of a compact extra
dimension with the size of the Planck length in order to formulate a
unification theory of electromagnetism and gravitation. More recently,
braneworld models, originally conceived as a framework to solve the hierarchy
problem, have vigorously renewed the interest in this subject
\cite{ADD:1,ADD:2,RS:1,RS:2}. In the braneworld scenario, our observable
universe is a submanifold isometrically embedded in a higher-dimensional space
and, unlike the Kaluza-Klein model, the size of the extra dimensions can be
much larger than Planck length or even infinite \cite{RS:2} without
introducing any irremediable conflict with experimental data.

The main feature of the braneworld scenario is that the standard model fields
are confined in a submanifold ($3$-brane), if they are not excited above a
certain energy scale,\ which should be greater than TeV order \cite{lhc}. On
the other hand gravity has access to extra dimensions. Despite this freedom,
due to the existence of a bound zero-mode, the gravitational force recovers
its four-dimensional behavior for distances much larger than the
characteristic length scale, $\ell$, of the supplementary space. In models
with compact dimensions, $\ell$ corresponds to the size of the extra
dimension, while in models of non-compactified extra dimension, $\ell$ is
related to the curvature radius of the ambient space. Below this length scale,
the influence of extra dimensions on the gravitational force become
significant. It happens that $\ell$ can be much greater than the scale,
$L_{m}$, in which the confined fields feel the effects of the extra dimension
directly. In the original ADD model, $L_{m}<10^{-19}%
\operatorname{m}%
$ \cite{ADD:1}, while, for branes inspired in string theories, $L_{m}$ is much
smaller \cite{ADD:2}.

In sight of this, the search for traces of extra dimensions in the
gravitational interaction arises as a paramount issue which can be empirically
investigated. Direct tests of the gravitational force in microscopic domains,
based on torsion balance experiments, found no deviation from the inverse
square law, constraining, in this way, the radius $R$ $\left(  =\ell
/2\pi\right)  $ of the extra dimension. Considering the original ADD model
\cite{ADD:1}, it is found that $R<44%
\operatorname{\mu m}%
$, which is the tightest bound for models with only one extra dimension
\cite{LONG,Hoyle:2001,Hoyle:2004,Hoyle:2007,review}. For greater codimensions,
the most stringent constraints come from astrophysics \cite{SN,neutronstar}
and high-energy particle collisions \cite{GRW,lhc}.

Here we study the effects of the extra dimensions in a bound system, in order
to find independent constraints for the number and the size of the
supplementary space. It is reasonable to expect that the extra dimensions
modify significantly the behavior of gravity in distances shorter than $\ell$.
If the gravitational field in the weak-field limit obeys Gauss law, then the
gravitational potential of a point-like mass $m$ has the following behavior:%

\begin{equation}
V\left(  r\right)  =\left\{
\begin{array}
[c]{l}%
-\frac{Gm}{r},\text{ for }r\gg\ell\\
-\frac{G_{n}m}{r^{n+1}},\text{ for }r\ll\ell
\end{array}
\right.  ,
\end{equation}
where $G_{n}$ is the gravitational constant in the higher-dimensional space
with $n$ extra dimensions. The exact relation between $G_{n}$ and the
Newtonian constant $G$ depends on the extra-dimensional volume, which, on its
turn, depends on the topology and curvature of the supplementary space.
However, in magnitude order, it is expected that $G_{n}\sim G\ell^{n}$.

In atomic systems the usual gravitational interaction is negligible. The
average potential gravitational energy of a hydrogen atom in the ground state
is of the order of $Gm_{p}m_{e}/a_{0}\sim10^{-38}%
\operatorname{eV}%
$, where $a_{0}$ is the Bohr radius. However, due to effects of the extra
dimensions, the potential energy increases by a factor of the order of
$\left(  \ell/r\right)  ^{n}$ for short distances ($r<\ell$). Thus, depending
on the values of $\ell$ and $n$, the gravitational energy could in principle
modify the energy spectrum of the hydrogen significantly. Therefore, the
highly accurate measurements of atomic transitions can be used to put
empirical constraints on models of extra dimensions.

Treating the gravitational potential as a perturbation of the Hamiltonian of
the hydrogen atom, we find that the energy shift caused by extra dimensions in
a certain quantum state $\psi$ is proportional to $\left\langle r^{-(n+1)}%
\right\rangle $, i.e., the average of $r^{-\left(  n+1\right)  }$ in that
state. It happens that, for $n\geq2$, this average diverges for $S$-levels.
This problem is connected with the fact that the short-distance behavior of
the brane-to-brane graviton propagator is not computable in the original ADD
model, even in the tree level, suggesting that it may depend on the
ultraviolet details of a fundamental quantum theory of gravitation, as it was
pointed out in Ref \cite{GRW}, where the effects of extra dimensions on
high-energy particle collision via the gravitational interaction were investigated.

The first attempts to constrain extra-dimensional parameters by using atomic
spectroscopy have tried to circumvent the divergence problem by establishing a
cut-off radius \cite{specforxdim1,atomicspec,Li,Wang,specforxdim2}. However,
proceeding in that way, the constraints become dependent on this arbitrary
parameter. Another way to avoid the divergence problem is by considering the
effects of the modified potential on molecular spectroscopy instead of atomic
spectroscopy \cite{molecule}. However, this approach gives much weaker constraints.

In the classical level, the divergence problem arises because the matter
distribution confined in the brane is singular from the perspective of the
ambient space. It happens that such delta-like confinement is just an
idealization that hides the internal structure of the brane. The relevance of
the brane substructure to regulate divergences of a thin brane effective
theory has been observed and studied before \cite{effectivebrane,thickeffect}.
In Ref \cite{effectivebrane}, for instance, these divergences are treated by
introducing a renormalization prescription. Here we follow a different
procedure which relies on the fact that, in the thick brane scenario, the wave
function of localized particles has a certain width $\sigma$ in the
extra-dimensional space. As we shall see, when the width is taken into
account, the average $\left\langle r^{-(n+1)}\right\rangle $ is finite and
depends on $\sigma$. As the value of width is bounded by the thickness of the
brane, then our analysis provides a joint constraint of the higher-dimensional
Planck mass and brane thickness.

\section{An atom in a thick brane}

In the context of string theories, the brane may be considered as infinitely
thin. However, in the framework of the field theories, it may have a
thickness. For instance, if we consider an ambient space of (4+1)-dimensions,
a 3-brane may be represented as a domain wall solution of some scalar field
\cite{rubakov}. A Yukawa-type interaction between Dirac spinor fields and the
scalar field can localize matter in the core of the domain wall. Due to this
interaction, the state of zero-mode of the spinor field is given by the
following wave function:%
\begin{equation}
\Psi\left(  \mathbf{x},z\right)  =\exp\left[  -h\int_{0}^{z}\phi_{0}\left(
y\right)  dy\right]  \psi\left(  \mathbf{x}\right)  , \label{wave}%
\end{equation}
where $h$ is the coupling constant, $\psi\left(  \mathbf{x}\right)  $
represents a free spinor in the $\left(  3+1\right)  $-dimension, $\phi_{0}=$
$\eta\tanh\left(  z/\varepsilon\right)  $ is the scalar in a domain wall
configuration interpolating between two vacua $\pm\eta$ of the scalar field.
With respect to the transversal direction $z$, the wave function has a peak at
$z=0$ and exponentially falls as we move away from the center of the domain
wall. The quantity $\varepsilon$ defines the thickness of the brane and must
be smaller than $10^{-19}%
\operatorname{m}%
$, according to current experimental constraints \cite{ADD:1,lhc}.

Following a similar reasoning, it is possible to devise a confinement
mechanism for matter in topological defects of greater codimensions. By
generalizing the above result for a space with $n$ extra dimensions, we can
consider that, from the phenomenological point of view, the wave function of
localized particles can be written as $\Psi\left(  \mathbf{x},\mathbf{z}%
\right)  =\chi\left(  \mathbf{z}\right)  \psi\left(  \mathbf{x}\right)  $. For
the sake of simplicity, we will assume that the extra-dimensional part $\chi$
is a Gaussian function around the center of the brane with a standard
deviation $\sigma$:%
\begin{equation}
\chi\left(  z\right)  =\left(  \frac{2}{\pi\sigma^{2}}\right)  ^{n/4}%
\exp\left(  -%
{\displaystyle\sum\limits_{i=1}^{n}}
\frac{z_{i}^{2}}{\sigma^{2}}\right)  . \label{extrapart}%
\end{equation}
Of course the width of the wave function in the transverse directions should
satisfy $\sigma<\varepsilon<\ell.$ At this point it is important to emphasize
that, although we have chosen a Gaussian profile in the transverse direction,
$\chi$ can be any normalizable function of $z$, as we shall see later.

Now we can describe the gravitational interaction between the proton and the
electron in this scenario. In the linear regime of gravity, there exists a
coordinate system (a gauge) in which the gravitational potential satisfies the
Laplace equation in vacuum \cite{myers}. Thus, in the first order of $G_{n}$,
the exact potential produce by a point-like mass in a higher-dimensional space
can be obtained, at least formally, for any compact topology \cite{kehagias}.
For instance, if the supplementary space has a topology of a $n-$dimensional
torus with size $\ell$, then the gravitational potential of a mass $m$ at the
point $\mathbf{R=}\left(  \mathbf{x},\mathbf{z}\right)  $ of the ambient space
is given by%
\begin{equation}
V\left(  \mathbf{R}\right)  =-\frac{G_{n}m}{R^{n+1}}-\sum_{i}\frac{G_{n}%
m}{\left\vert \mathbf{R}-\mathbf{R}_{i}^{\prime}\right\vert ^{n+1}}, \label{V}%
\end{equation}
where $\mathbf{R}_{i}^{\prime}=\ell\left(  0,0,0_{,}k_{1}...,k_{n}\right)  $
and each $k_{i}$ is an integer number. The potential (\ref{V}) is a solution
of the ($3+n$)-dimensional Laplace equation with the appropriate boundary
conditions. Indeed, the form of vector $\mathbf{R}_{i}$ ensures that $V\left(
\mathbf{R}\right)  $ is periodic with respect to the extra-dimensional
directions as required by the topology of the supplementary space. It is
interesting to note that, from the perspective of the space $%
\mathbb{R}
^{3+n}$, $V\left(  \mathbf{R}\right)  $ can be viewed as a superposition of
potentials generated by the real mass $m$ at the origin and by
copies\ (mirrors images) of the mass localized by the vectors $\mathbf{R}%
_{i}^{\prime}$.

In the far zone, i.e., for $\left\vert \mathbf{x}\right\vert >>\ell$,
(\ref{V}) reproduces the Newtonian potential. Moreover the corrections due to
the extra dimensions can be written as $1/r\left(  1+\alpha e^{-r/\lambda
}\right)  $ \cite{kehagias}. This Yukawa-law form is vastly employed to
constrain parameters of higher-dimensional models, by using data from the
torsion balance experiments \cite{review}. On the other hand, in a bound
system, such as the atom, the major contribution comes from the first term of
(\ref{V}), at least if the atom is at the lowest states. The first corrections
come from the mirror images closest to $m$. The potential of each one of these
$2n$ first neighbors is less than $G_{n}m/\ell^{n+1}$ $\sim Gm/\ell$, which is
negligible in comparison to the first term, as we can verify by calculating
this term with the extra-dimensional estimated size, $\ell$, that we obtain in
Figure 2. For a detailed estimate of the contribution of the mirror images
see, for instance, the appendix of Ref. \cite{protonpuzzle}.

Thus, for our purpose, we can approximate the gravitational potential of a
point-like source by the function $-G_{n}m/R^{n+1}$, which is proportional to
the Green function of the Laplace operator in the flat higher-dimensional
space $%
\mathbb{R}
^{n+3}$. Therefore, in this approximation, the potential produced by the
proton, assuming that its mass $m_{p}$ is distributed on the spatial extension
of the nucleus, is given by%
\begin{equation}
\phi\left(  \mathbf{R}\right)  =-G_{n}%
{\displaystyle\idotsint}
\frac{\rho\left(  \mathbf{R}^{\prime}\right)  }{\left\vert \mathbf{R}%
-\mathbf{R}^{\prime}\right\vert ^{n+1}}d^{3+n}\mathbf{R}^{\prime},
\label{potential}%
\end{equation}
where the mass density is $\rho=\left\vert \Psi_{p}\right\vert ^{2}m_{p}$ and
$\Psi_{p}\left(  \mathbf{x},\mathbf{z}\right)  =\chi\left(  \mathbf{z}\right)
\psi_{p}\left(  \mathbf{x}\right)  $ is the higher-dimensional wave function
of the proton. We assume that the mass is uniformly distributed inside the
nucleus, so the 3-dimensional part $\psi_{p}\left(  \mathbf{x}\right)  $ is
constant in the spatial extension of the nucleus and zero outside.

In the hydrogen the average kinetic energy of the electron is a small fraction
of its rest energy. Thus, in a first approach, we are going to consider the
non-relativistic picture of the atomic system. In the classical framework, the
gravitational interaction between electron and proton is governed by the
Hamiltonian $H_{G}=m_{e}\phi$. Now assuming that $H_{G}$ is a small term of
the hydrogen Hamiltonian, it follows, from the perturbation method, that the
energy shift corresponding to a certain atomic state, in the first order, is
given by%
\begin{equation}
\delta E_{\psi}=%
{\displaystyle\idotsint}
\left\vert \Psi_{e}\right\vert ^{2}m_{e}\phi\left(  \mathbf{R}\right)
d^{3+n}\mathbf{R}, \label{energy}%
\end{equation}
where $\Psi_{e}\left(  \mathbf{x},\mathbf{z}\right)  $ is the
higher-dimensional wave function of the electron (more precisely, the reduced
particle) which comprises the extra-dimensional part $\chi\left(
\mathbf{z}\right)  $ and the usual solutions $\psi_{e}\left(  \mathbf{x}%
\right)  $ of the Schr\"{o}dinger equation for the hydrogen atom.

Therefore, the extra dimensions modify the energy spectrum of the hydrogen by
means of the gravitational interaction between electron and proton.
Considering the wave functions of $1S$ and $2S$ states ($\psi_{1S}=1/\sqrt{\pi
a_{0}^{3}}\exp\left(  -r/a_{0}\right)  $ and $\psi_{2S}=1/\sqrt{8\pi a_{0}%
^{3}}\left(  1-r/2a_{0}\right)  \exp\left(  -r/2a_{0}\right)  $,
respectively), the shift in the frequency of the $2S-1S$ transition,
$\Delta\nu_{G}=\left\vert \delta E_{2S}-\delta E_{1S}\right\vert /h$, can be
calculated from (\ref{energy}) and compared with the experimental value.

\section{Results and Discussion}

The empirical value of the $2S-1S$ transition frequency in the hydrogen atom
is $f_{\exp}=2466061413187035%
\operatorname{Hz}%
$ with an experimental error $\delta_{\exp}=10%
\operatorname{Hz}%
$ \cite{H}. The theoretical prediction (based on well-established
four-dimensional physics) agrees with the experimental value up to the order
of the theoretical error ($\delta_{th}$), which corresponds to $32$ $%
\operatorname{kHz}%
$ and is related to uncertainties in measurements of the proton radius
\cite{pachuki}. Thus, to be consistent, any effect of extra dimensions should
be lesser than the combined errors. The constraints on the parameters of the
extra dimensional theory are obtained by imposing that $\Delta\nu_{G}%
<\sqrt{\delta_{th}^{2}+\delta_{\exp}^{2}}$. The numerical analysis of this
condition is summarized in Figure 1. It establishes a lower bound for the
fundamental Planck mass of the higher dimensional space $M_{D}$ in terms of
$\sigma$. According to Refs. \cite{GRW,myers}, the fundamental Planck mass is
related to $G_{n}$ by the formula $M_{D}^{2+n}=\Omega_{n}\left(
\hbar/c\right)  ^{n}\hbar c/G_{n}$, where $\Omega_{n}=\left(  2\pi\right)
^{n}\Gamma\left(  \frac{n+3}{2}\right)  /[(n+2)2\pi^{\left(  n+3\right)  /2}%
]$. In Figure 1, the regions below the curves are excluded. It is worth
mentioning that the width of the particle wave function in the transverse
directions, $\sigma$, is smaller than the thickness of the brane
($\varepsilon$). Therefore, the analysis shows that the constraints are
tighter for thinner branes.

The cases $n=1$ and $n=2$ are not shown because the corresponding constraints
are very weak. For $n\geq3$, the major contribution of the gravitational
energy comes from the integral in the interior of the nucleus, when we
consider realistic branes with $\varepsilon<10^{-19}%
\operatorname{m}%
$. The leading term is of the order of $G_{n}m_{p}m_{e}/a_{0}^{3}\sigma^{n-2}%
$. This dependence on $\sigma$ explains why narrow branes provide very
stringent limits on the fundamental Planck mass.%

\begin{center}
\includegraphics[
height=2.7259in,
width=2.9179in
]%
{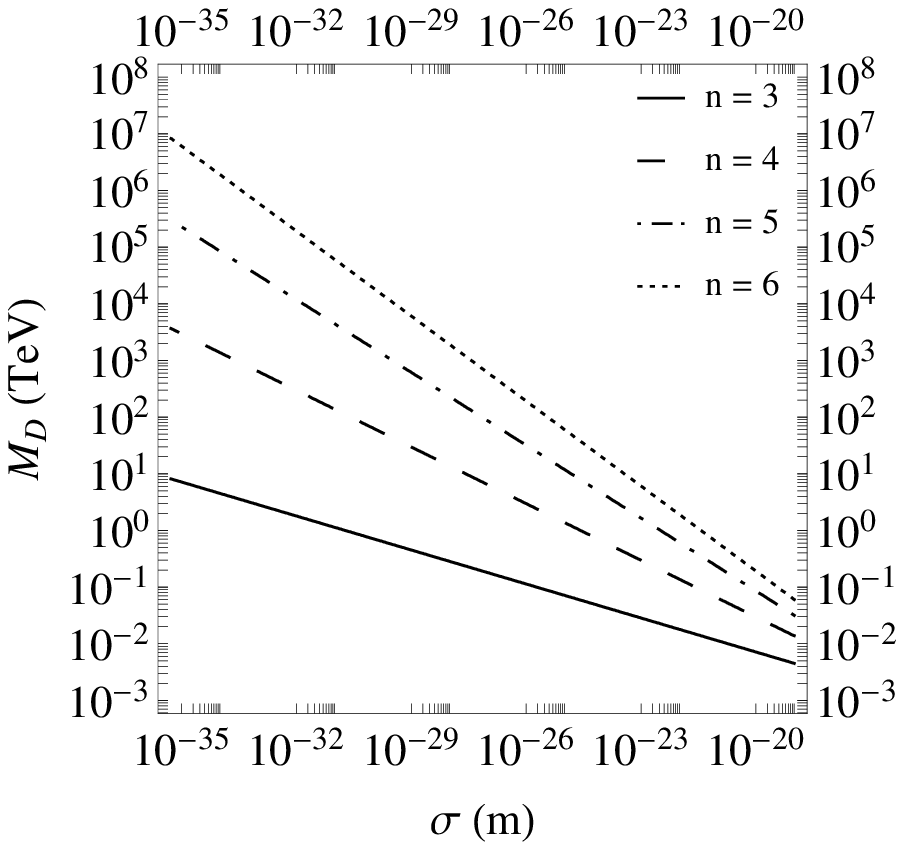}%
\\
Figure 1. {\protect\small Lower bound for the higher-dimensional Planck mass
M}$_{D}${\protect\small (expressed in natural units) in terms of }$\sigma
${\protect\small  (the wave function width in the transverse directions)}%
$.${\protect\small  The areas below the lines are excluded.}%
\label{Figure1}%
\end{center}

In the original ADD model, the supplementary space is a flat $n-$torus with
radius $R$. The relation between $R$ and the higher-dimensional Planck mass
$M_{D}$ (given by $G^{-1}=8\pi R^{n}\mathcal{M}_{D}^{2+n}$, where$\mathcal{M}%
_{D}^{2+n}=M_{D}^{2+n}/[\left(  \hbar/c\right)  ^{n}\hbar c]$, see \cite{GRW})
can be used to constraint the radius of the extra dimension. Figure 2 shows
upper limits on the extra dimension radius for $n=3,4,5$ and $6$. The regions
above the curves are excluded.%
\begin{center}
\includegraphics[
height=2.6351in,
width=2.9179in
]%
{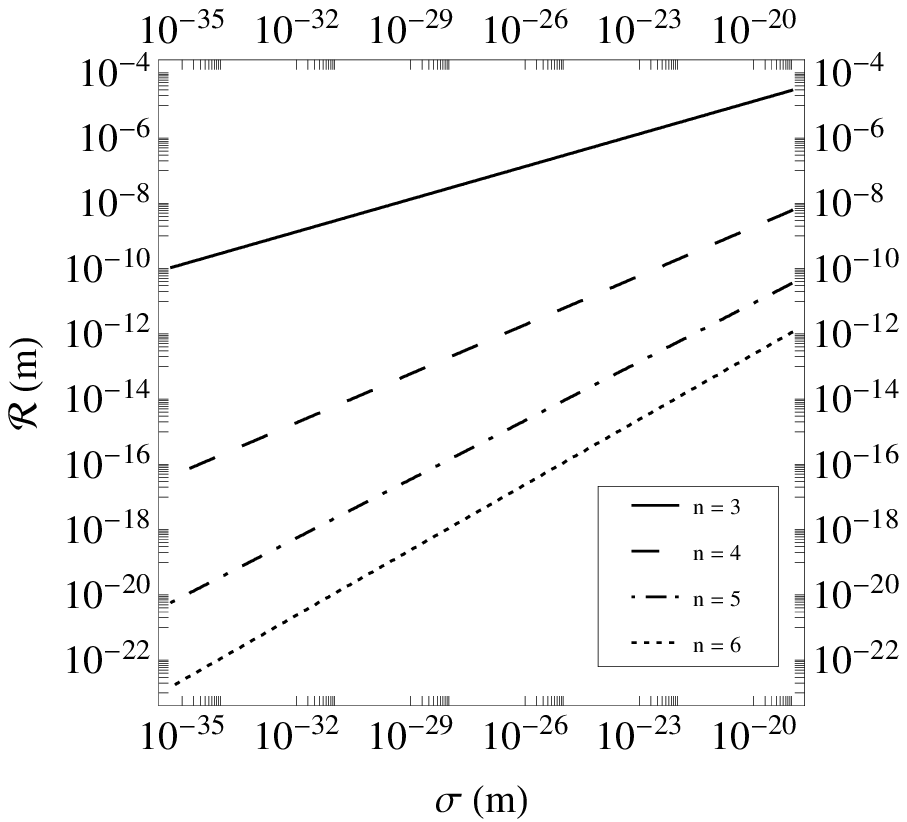}%
\\
{\protect\small Figure 2. Upper limit for the radius of the extra dimensions
in terms of }$\sigma${\protect\small  for n=3, 4, 5 and 6. The regions above
the lines are excluded.}%
\label{Figure2}%
\end{center}

As mentioned before, we have admitted that the wave function in the
supplementary space has a Gaussian profile (\ref{extrapart}) for the sake of
simplicity. Nevertheless, it is important to emphasize that this assumption
does not play an essential role in our analysis. Actually, our results do not
change, even if we consider other profiles, provided that the parameter
$\sigma$ is defined by:%
\begin{equation}
\frac{1}{\sigma^{m}}\equiv\frac{\Gamma\left(  n/2\right)  }{\Gamma\left(
\frac{n-m}{2}\right)  }\int\frac{\left\vert \chi_{p}\left(  z_{1}\right)
\right\vert ^{2}\left\vert \chi_{e}\left(  z_{2}\right)  \right\vert ^{2}%
}{\left\vert \vec{z}_{1}-\vec{z}_{2}\right\vert ^{m}}d^{n}z_{1}d^{n}z_{2},
\end{equation}
where $m$ is a positive integer that should satisfy the condition
$m\leq\left(  n-1\right)  $ and $\Gamma$ stands for gamma function. When the
profile is Gaussian, this parameter $\sigma$ corresponds to the width of the
Gaussian function.

At this point, we should stress that the above constraints were derived based
on the classical regime of the gravitational interaction. Nevertheless, as it
was emphasized in Ref. \cite{GRW}, quantum-gravity effects may become relevant
in a length scale of the order of $l_{D}$ (the higher dimensional Planck
length, given by $l_{D}=\left(  \hbar/c\right)  M_{D}^{-1}$ ) or even in a
greater scale, depending on the quantum-gravity theory, not yet known. If this
is the case, then unexpected effects may emerge and could distort or even
overshadow the classical results we have investigated here.

In the lack of a quantum-gravity theory, let us explore some possibilities.
According to \cite{effgravity}, quantum corrections to the gravitational
potential energy may be estimated by treating General Relativity theory as an
effective theory. The classical potential energy, in three-dimensional space,
is given by the Newtonian term $GMm/d$, where $d$ is the distance between
particles with mass $M$ and $m$. The quantum contributions are smaller by a
factor of the order of $\left(  l_{p}/d\right)  ^{2}$, where $l_{p}$ is the
Planck length of the ordinary three-dimensional space \cite{effgravity}. In
the higher-dimensional case, the classical potential is $G_{n}Mm/d^{n+1}$ and
the quantum corrections would be of the order of $\left(  l_{D}/d\right)
^{n+2}$, based on dimensional analysis. Now, considering the hydrogen atom, it
is instructive to define an effective extra-dimensional distance, $d_{eff}$,
between proton and electron, though they cannot be considered as point-like
particles in this bound system. By writing the atomic gravitational energy as
$G_{n}m_{p}m_{e}/d_{eff}^{n+1}$, it follows that $d_{eff}^{n+1}\sim
\sigma^{n-2}a_{0}^{3}$ in the ground state. Notice that $d_{eff}$ is a kind of
geometric average of the wave packet widths in the transversal and parallel
direction of the brane. Now, from the lower constraints on $M_{D}$, we can
estimate upper bounds for $l_{D}$ as a function of $\sigma$. By a direct
calculation, we can verify that the ratio $\left(  l_{D}/d_{eff}\right)
^{n+2}$ depends on $n$ and $\sigma$, but, for any dimension and for any value
of $\sigma$ investigated here, it is smaller than $10^{-8}$. Thus, if
$d_{eff}$ is the relevant characteristic length scale that regulates the
gravity behavior in this system, then we may expect that the classical
contribution will be the leading gravitational influence in this context.
However, as the fundamental quantum-gravity theory is not known, only
experiments can decide on this question.

\section{Comparison with other constraints}

Experimental bounds for the fundamental Planck mass are determined from many
areas of Physics. As we have mentioned before, when the codimension is greater
than 2, some of the most stringent constraints of $M_{D}$ are established by
high-energy particles collisions. Data analysis from monojet events in
proton-proton collisions at the LHC gives the following lower bounds for
$M_{D}$ in TeV/c$^{2}$: $4.38$ $\left(  n=3\right)  $, $3.86$ $\left(
n=4\right)  ,$ $3.55$ $\left(  n=5\right)  $ and $3.26$ $\left(  n=6\right)  $
\cite{monojet,landsberg}. Examining Figure 1, we may say that, in this thick
brane scenario, unless unpredictable quantum-gravity effects suppress the
classical result, spectroscopy could give stronger constraints in the case the
confinement parameter, $\sigma$, is small.

However, as $\sigma$ has an unknown value, it is interesting to investigate
the inverse problem, i.e., to estimate the maximum influence of the
gravitational energy on the hydrogen spectrum, considering the current
constraints on $M_{D}$ given by the LHC. In Figure 3, we show the
gravitational contribution to the hydrogen $\left(  2S-1S\right)  $
transition, $\Delta E_{G}$, as a function of the confinement parameter,
$\sigma$.%
\begin{center}
\includegraphics[
height=2.6264in,
width=2.9179in
]%
{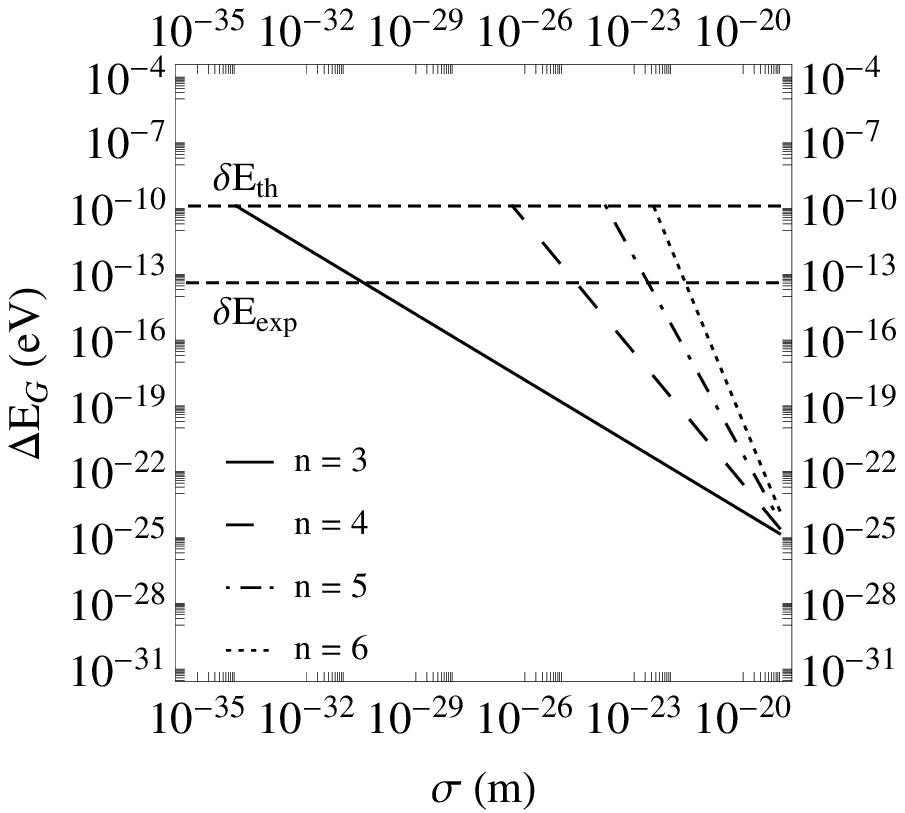}%
\\
Figure 3. {\protect\small The energy gap between the }$2S-1S$
{\protect\small states in the hydrogen atom due to higher-dimensional
gravitational interaction, assuming LHC constraints for} $M_{D}$%
{\protect\small . The horizontal lines, labeled with }$\delta E_{\exp}$
{\protect\small and} $\delta E_{th}${\protect\small  are the current
experimental precision and theoretical uncertainty, respectively, in the
}$2S-1S$ {\protect\small transition.}%
\label{Figure3}%
\end{center}

Here the gravitational energy $\Delta E_{G}$ is confronted with the current
theoretical uncertainty $\delta E_{th}$ and current experimental error $\delta
E_{\exp}$. For each value of $n$ considered, there is an interval in the
$\sigma$-axis in which $\Delta E_{G}$ is less than the theoretical
uncertainty. However, even in this range, extra dimensions induce a huge
amplification of the atomic gravitational energy in comparison with the
ordinary three-dimensional case, which is of the order of $10^{-38}$ $%
\operatorname{eV}%
$. Within these specific ranges of $\sigma$, Figure 3 indicates how much the
experimental precision should be improved in order to the hydrogen
spectroscopy become capable of revealing traces of the supposed extra
dimensions. For shorter $\sigma$, the limit on $\Delta E_{G}$ is still given
by $\delta E_{th}$. Once again, it is important to emphasize that this
prediction is valid only if no quantum-gravity effects suppress the classical
result. But, notice that, if quantum-gravity effects reinforce the classical
contribution, then signals of extra dimensions, though unpredictable, would be
viewed earlier.

New forms of constraints can be obtained by considering relations between
$\sigma$ and $l_{D}$ in advance. Let us admit that $l_{D}$ is smaller than
$\sigma$ by a factor $x$, i.e., $l_{D}=x\sigma$, with $x<1$. In this case, the
condition that the energy shift between $2S$ and $1S$ states, due to the
gravitational interaction, be smaller than the uncertainty $\delta E$ yields
the following constraint:%
\begin{equation}
c^{2}M_{D}>\left[  \alpha_{n}\left(  \frac{\hbar c}{a_{0}}\right)  ^{3}%
\frac{m_{p}c^{2}m_{e}c^{2}}{\delta E}\right]  ^{1/4}x^{\left(  n-2\right)
/4}, \label{MDx}%
\end{equation}
where $\alpha_{n}=7\gamma_{n}\Omega_{n}/8\pi$ and coefficients $\gamma_{n}$
have the following values: $\gamma_{3}=2\pi^{3/2},\gamma_{4}=4\pi/3,\gamma
_{5}=\pi^{3/2}/3$ and $\gamma_{6}=$ $4\pi/15$. Thus, for instance, if
$\sigma=10l_{D}$, then the lower bounds for $M_{D}$, assuming $\delta E$ of
about $10^{-10}$ $%
\operatorname{eV}%
$, would be of GeV order, therefore, $10^{3}$ weaker than the constraints from
LHC. Equation (\ref{MDx}) also shows a weak sensitivity of $M_{D}$ with
respect to the experimental precision. Indeed, if $\delta E$ was reduced by a
factor of $10^{4}$, then, $M_{D}$ would be improved $10$ times, only. It is
also interesting to compare these constraints with indirect limits obtained in
lepton colliders, which seem to have a similar origin. Ref. \cite{GRW}
determined the maximum $M_{D}$ sensitivity which can be reached by studying
the final state with photons and missing energy at an electron-positron
collider. Considering $\sqrt{s}=1$ TeV and integrated luminosity
$\mathcal{L}=200$ fb$^{-1}$, the predicted bounds for $M_{D}c^{2}$ in TeV are:
$4.0$ $\left(  n=3\right)  $, $3.0$ $\left(  n=4\right)  ,$ $2.4$ $\left(
n=5\right)  $, when the beam polarization is 90\%. These limits are much
greater than those from equation (\ref{MDx}).

Another way to find new constraints is, instead of fixing a direct relation
between $l_{D}$ and $\sigma$, to require that the proton energy density in the
extra dimensional space, $\rho$, be smaller than the critical density,
$\rho_{D}=M_{D}c^{2}/l_{D}^{n+3}$, by some factor $y.$ Admitting a Gaussian
profile in the transverse direction, we can estimate the proton density, $\sim
m_{p}c^{2}/\sigma^{n}R_{p}^{3}$, in the center of the thick brane. From the
condition $\rho/\rho_{D}=y$, we can write $\sigma$ in terms of $y$. By using
this relation, we find constraints for the fundamental Planck mass in terms of
$y$:%
\begin{equation}
c^{2}M_{D}>\beta_{n}\left(  \frac{m_{p}c^{2}}{\left(  R_{p}^{3}/\hbar^{3}%
c^{3}\right)  }\right)  ^{1/4}\left(  \frac{R_{p}^{3}m_{e}c^{2}}{a_{0}%
^{3}\delta E}\right)  ^{n/8}y^{\left(  n-2\right)  /8}, \label{MDy}%
\end{equation}
where $\beta_{n}=\alpha_{n}^{n/8}\left[  \left(  4\pi/3\right)  \left(
\pi/2\right)  ^{n/2}\right]  ^{\left(  n-2\right)  /8}$. Thus, taking $y=0.1$,
to make some estimation, we find $M_{D}>1.6$ GeV/c$^{2}$ $\left(  n=3\right)
$ to $M_{D}>76$ GeV/c$^{2}$ $\left(  n=6\right)  $, considering the current
theoretical uncertainty $\delta E_{th}$. In comparison with the previous case,
the constraints are much more sensitive with respect the precision $\delta E$.
Indeed, if the condition (\ref{MDy}) is calculated with the current
experimental error, $\delta E_{\exp}$, the constraints will be at least
$10^{3n/8}$ times stronger. Specifically, for $n=6$, we would have $M_{D}>$
$32$ TeV/c$^{2}$, which is more stringent than the collider constraints.

\section{Final remarks}

In the thick brane scenario, it is expected that the matter and the standard
model fields do not feel directly the effects of the extra dimension in a
length scale less than the thickness of the brane $\varepsilon$, while the
modification in the gravitational field may arise in a scale $\ell$ much
greater than $\varepsilon$. Taking this into account, we have seen that the
corrections of the gravitational potential due to extra dimensions may give
significant contribution in a bound system. Thus, by using precision
measurements of $2S-1S$ transition frequency of the hydrogen, we have obtained
new constraints for the higher-dimensional Planck mass in terms of a
confinement parameter, $\sigma$, of the matter inside the brane, which should
be smaller than the brane thickness.

The constraints we find here are stronger for thinner branes. In fact,
comparing our results with the current constraints from LHC data, which go
from $M_{D}>4.38$ TeV/c$^{2}$ $\left(  n=3\right)  $ to $M_{D}>3.26$
TeV/c$^{2}$ $\left(  n=6\right)  $ \cite{monojet,landsberg}, we may conclude
that, for some narrow branes, atomic spectroscopy could impose very stringent
constraints for the fundamental Planck mass in higher-dimensional spaces.

Finally, it is important to emphasize that, as the higher-dimensional Planck
mass can be much smaller than the four-dimensional Planck scale, it is
expected that quantum-gravity effects may arise much sooner in comparison with
the traditional picture without extra dimensions \cite{GRW}. However, as the
fundamental theory is not known, the supposed quantum effects are
unpredictable. On the other hand, the constraints from the hydrogen
spectroscopy we find here relies on the classical behavior of gravity, thus,
it is important to highlight that the present results are validity only if the
classical contributions are not suppressed by quantum-gravity effects.

\subsection{Acknowledgment}

A. S. Lemos thanks CAPES for financial support.

\end{document}